\documentclass[12pt]{article}
\usepackage{graphicx}
\begin{document}

\title{
 Thermodynamics of k-essence}
\author{
Neven Bili\'c\thanks{E-mail: bilic@thphys.irb.hr}
 \\
 Rudjer Bo\v{s}kovi\'{c} Institute, 10002 Zagreb, Croatia 
}
\maketitle

\begin{abstract}
We discuss thermodynamic properties of 
dark energy using the formalism of 
field theory at finite
temperature. In particular, we apply our formalism to a purely kinetic
type of k-essence.
We show quite generally that the entropy associated with dark energy is
always equal or greater than zero. Hence, contrary to often stated
claims, a violation of the null energy condition (phantom dark energy)
does not necessarily yield a negative entropy. In addition, we find that
the thermal fluctuations of a k-essence field may be represented by a free boson 
gas with an effective number of degrees of freedom equal to $c_s^{-3}$.
\end{abstract}



\section{Introduction}
In a number of recent papers
\cite{bre,lim,gonz,izq,moh,set,wan,gong,san,wan2,per,die,set2},
thermal properties of dark energy
have been discussed based on the assumption that
the dark energy substance is a thermalized
ensemble at a certain temperature
with an associated  thermodynamical entropy.
It is usually assumed that this temperature is an intrinsic property
of dark energy (DE) rather than the temperature of the heat bath
fixed by surrounding matter. 
A popular model of DE is the so-called {\em k-essence} 
\cite{arm2} which was originally introduced as a model for inflation \cite{arm1}.
The purpose of this paper is 
to analyze thermal properties of  a grand canonical system
described by a purely kinetic k-essence at nonzero temperature.
Our aim  is to extend and analyze classical solutions to field equations
at nonzero temperature.

Dark energy  is usually assumed to be barotropic, i.e.,
 described by an equation of state (EOS)
 in the form $p=p(\rho)$. Equivalently, one defines a field theory
Lagrangian, in which DE is described in terms of
 a classical self-interacting field
coupled to gravity (for a recent review, see \cite{cop}).
 Then, the EOS may  be deduced from the
energy-momentum tensor  obtained from the variational principle.
In order to explain an accelerated expansion 
the DE fluid must violate the {\em strong energy condition} \cite{ell,viss},
which requires  $3p+\rho \geq 0$ together with 
$p+\rho \geq 0$.
The so-called {\em phantom} models of DE have $p+\rho <0$,
thus  violating even the {\em null energy condition}.

Clearly, from the EOS alone
it is not possible to uniquely determine
the thermodynamic properties of a system.
One simple example is the EOS
$p=\rho/3$ which may describe a massless boson gas at $T\neq 0$ (hence $S\neq 0$)
but also a massless degenerate Fermi gas at $T=0$ (hence $S=0$).
A similar situation occurs for any barotropic EOS.

A consistent grand canonical description of DE
involves the thermodynamic equations with two variables:
the temperature $T$ and chemical potential $\mu$.
The chemical potential is associated
with a conserved particle number  $N$
related to the shift symmetry
$\theta \rightarrow \theta + {\rm const}$.
As we will demonstrate, the resulting thermodynamic equations do not require negative
entropy even in the phantom regime when the null energy condition is violated.
We show that if there exist a nontrivial, stable configuration
which we call {\em condensate}
characterized by the pressure $p_{\rm cd}$ and the density $\rho_{\rm cd}$,
then the $p_{\rm cd}+\rho_{\rm cd}$ term in the expression for entropy
is precisely canceled out by
the particle-number term.  The  only nonvanishing
contribution to the entropy is due to thermal fluctuations
that yield a thermal ensemble similar to a boson gas at nonzero temperature.

We organize the paper as follows: In section \ref{pure} we recapitulate
the basic hydrodynamics of
purely kinetic k-essence. In section \ref{chemical} we introduce 
the chemical potential
corresponding to the
conserved charge which is related to the shift symmetry.
Basic thermodynamics is discussed in section \ref{thermodynamic}.
The grand canonical and canonical partition functions are derived
in section \ref{thermal} where we also include a brief comment
concerning a general k-essence.
 Discussion and conclusions are given in section
\ref{conclusion}.


\section{Purely kinetic k-essence}
\label{pure}
Consider the action
\begin{equation}
S = \int \, d^{4}x \, \sqrt{- g}  \left[ - \frac{R}{16\pi} + {\cal L} (X) \right],
\label{eq4001}
\end{equation}
where $R$ is the curvature scalar and
\begin{equation}
{\cal L} = m^4 W(X)\, ;\; \;\;\;\; X \equiv g^{\mu \nu} \theta_{, \mu} \theta_{, \nu}
\label{eq1203}
\end{equation}
is the Lagrangian for
 the scalar field 
$\theta$ of dimension $m^{-1}$. The dimensionless function $W$
depends only on the dimensionless quantity $X$.
Such theories have been exploited as models for inflation
and dark 
matter/energy, for example a purely kinetic k-essence \cite{arm2,arm1,chi,sch1}
or ghost condensate \cite{ark1,kro1,kra1}.
The simplest nontrivial example of purely kinetic k-essence
is a ghost condensate Lagrangian \cite{ark1} 
\begin{equation}
{\cal L}_{\rm gh}=m^4 (1-X)^2 ,
\label{eq0071}
\end{equation}
which has been also studied in the context of dark matter/energy
unification
\cite{sch1}.
Another example originates from
the Dirac-Born-Infeld description of a $d$-brane in string theory
with the scalar Born-Infeld type Lagrangian
\begin{equation}
{\cal L}_{\rm DBI}=-m^4\sqrt{1-X}\, .
\label{eq0070}
\end{equation}
This Lagrangian 
was derived  from the Nambu-Goto action for a $d$-brane moving in the
$d$+2-dimensional bulk \cite{bor,jac} (for a simple derivation, see also
\cite{bil7}).
It may be easily seen that (\ref{eq0070}) yields
the EOS
$p\propto -\rho^{-1}$
of the Chaplygin gas,
 an exotic fluid 
 which has been suggested as a model for
unification of dark energy and dark matter
\cite{kam,bil4,fab}.
The generalization to
$p\propto -\rho^{-\alpha}$ ($0 \leq \alpha \leq 1$),
was suggested \cite{bent6} and shown to derive from
the Lagrangian
\begin{equation}
{\cal L}_{\rm gen}=-m^4\left(1-X^{(1+\alpha )/2\alpha}\right)^{\alpha/(1+\alpha )}
\label{eq007}
\end{equation}
which represents yet another example of purely kinetic k-essence.
  Subsequently,
 the term ``quart\-essence" was invented \cite{makl7}
to denote unified dark matter/energy models.

From (\ref{eq1203})
the equation of motion for $\theta$ follows
\begin{equation}
({\cal L}_X g^{\mu \nu} \theta_{, \nu})_{;\mu}=0 ,
\label{eq1304}
\end{equation} 
where ${\cal L}_X $ denotes the partial derivative with respect to $X$.
Equation (\ref{eq1304}) 
implies the existence of  a  conserved
current
\begin{equation}
j^{\mu}=\frac{2}{m} {\cal L}_X g^{\mu\nu}\theta_{,\nu}\, .
\label{eq3001}
\end{equation}
related to the invariance under the constant shift 
 $\theta\rightarrow \theta+{\rm const}$ of the scalar field $\theta$.
The shift symmetry reflects a correspondence between a purely kinetic k-essence
and a U(1) symmetric complex scalar field theory
in the so-called Thomas-Fermi approximation
\cite{bil4,par}.
 It may be shown \cite{bil6} that 
the Lagrangian (\ref{eq1203})
is equivalent to 
\begin{equation}
{\cal{L}}
 = \eta g^{\mu \nu} {\Phi^*}_{, \mu}
\Phi_{, \nu}
 -m^4 U(\eta|\Phi|^{2}/m^2),
\label{eq1101}
\end{equation}
if the amplitude  of the complex scalar field 
$\Phi=|\Phi| e^{im\theta}$ varies sufficiently slowly.
Here $\eta=1$ for a canonical scalar field
and $\eta=-1$ for a phantom.
The potential $U(Y)$ is related to $W(X)$ 
in (\ref{eq1203}) by a Legendre transformation
\begin{equation}
W(X)+U(Y)=
   XY ,
  \label{eq1105}
\end{equation}
with $X$ and $Y$ satisfying 
\begin{equation}
X= \frac{dU}{dY}\, ,
 \label{eq1406}
\end{equation}
\begin{equation}
Y= \frac{dW}{dX}\, .
 \label{eq1106}
\end{equation}
Then the current (\ref{eq3001}) corresponds
to the Klein-Gordon current
\begin{equation}
j_{\rm KG}^{\mu}=ig^{\mu\nu}(\Phi^*\Phi_{,\nu}-\Phi\Phi^*_{,\nu}),
\label{eq3004}
\end{equation}
and hence, 
 the conserved quantity $N$ corresponds to the usual $U(1)$ charge
of the complex scalar field.

Assuming $X>0$ 
the field $\theta$ may be regarded as a
velocity potential for the fluid 4-velocity
\begin{equation}
u^{\mu} = g^{\mu \nu} \theta_{,\nu} / \sqrt{X} 
\label{eq1215}
\end{equation}
satisfying the normalization condition
$u_{\mu} u^{\mu}$ = 1.
As a consequence,
 the energy-momentum tensor  
\begin{equation}
T_{\mu\nu}= 2{\cal L}_X
\theta_{,\mu}\theta_{,\nu}
-{\cal L}g_{\mu\nu}
\label{eq509}
\end{equation}
derived
from the Lagrangian ${\cal L}$ in (\ref{eq4001})
takes the perfect fluid
form, 
\begin{equation}
T_{\mu\nu}= (\rho+p) u_\mu u_\nu - p g_{\mu\nu}\, ,
\label{eq510}
\end{equation}
with the parametric equation
of state
\begin{equation}
p ={\cal L}\, ,
\label{eq4003}
\end{equation}
\begin{equation}
\rho = 2 X {\cal L}_{X}-{\cal L}\, ,
\label{eq4004}
\end{equation}
and the speed of sound
\begin{equation}
c_{s}^{2} = \frac{p_X}{\rho_X}.
\label{eq4010}
\end{equation}
As before, the subscript $X$ denotes the partial derivative with respect
to $X$.
A perfect fluid description applies only for $X>0$.
Furthermore, equations (\ref{eq4003}) and (\ref{eq4004}) imply
 that the domains where
${\cal L}_X>0$ correspond to a canonical scalar field 
($p+\rho>0$)
and those where ${\cal L}_X<0$ to a phantom.
In particular, if in the neighborhood of $X=0$,
${\cal L}\sim \eta X$,  the kinetic term
is  canonical for $\eta=1$
and is of phantom type for $\eta=-1$.
The field 
that behaves as phantom near $X=0$
 is also known under the name {\em ghost}.

The conserved particle number associated with the current
(\ref{eq3001})
 is 
\begin{equation}
N=\int_{\Sigma} j^\mu d\Sigma_{\mu}
=\int_{\Sigma} n\, u^{\mu}d\Sigma_{\mu}\, ,
\label{eq3002}
\end{equation}
where the integration goes over an arbitrary spacelike hyper-surface $\Sigma$
 that contains
the ``particles''.
Using the definition (\ref{eq1215}) for the velocity, we obtain the 
particle-number density 
\begin{equation}
n= \frac{2}{m}\sqrt{X} {\cal L}_X \, .
\label{eq3103}
\end{equation}
A similar expression,
with the right-hand side differing only in a dimensionfull constant factor,
 has been 
derived also in \cite{die}.

\section{Effective Lagrangian}
\label{chemical}
In this section we construct the effective Lagrangian 
as a function of the chemical potential $\mu$ associated with the conserved
particle-number (\ref{eq3002}).
In the Hamiltonian formulation \cite{wal} we choose
the hyper surface $\Sigma$ at constant time so that the total
number of particles
(\ref{eq3002}) becomes a volume integral
\begin{equation}
N= \frac{2}{m} \int_V  {\cal L}_X g^{0\nu}\theta_{,\nu}dV
=\frac{1}{m}\int_V  \frac{\partial {\cal L}}{\partial\theta_{,0}} dV .
\label{eq3003}
\end{equation}
Next, we define the grand canonical partition function
\begin{equation}
Z=\mbox{Tr}\; e^{-\beta (\hat{H}-\mu \hat{N})}=\int [d\pi]
\int_{\rm periodic} [d\theta]\exp \int_0^\beta d\tau \int dV \left( i\pi 
\frac{\partial\theta}{\partial\tau}
-{\cal H}(\pi,\theta_{,i}) +\frac{\mu}{m}\pi\right),
\label{eq3005}
\end{equation}
where $\tau$ denotes Euclidean time. The Lorentzian
and  Euclidean times  are related by 
\begin{equation}
t=-i\tau ,
\label{eq3105}
\end{equation}
as usual.
In equation (\ref{eq3005}) we employed (\ref{eq3003}) to express the particle number $N$
in terms of the conjugate momentum field 
\begin{equation}
\pi=\frac{\partial {\cal L}}{\partial\theta_{,0}}\, .
\label{eq3007}
\end{equation}
The Hamiltonian  density $\cal H$ is defined as usual by the Legendre transformation
\begin{equation}
{\cal H}(\pi, \theta_{,i})= \pi\theta_{,0}-{\cal L}(\theta_{,0},\theta_{,i})\, ,
\label{eq3006}
\end{equation}
with
\begin{equation}
\theta_{,0}=\frac{\partial {\cal H}}{\partial\pi}\, .
\label{eq3008}
\end{equation}
A formal functional integration of (\ref{eq3005}) over $\pi$ yields the partition function 
expressed in terms of the effective Euclidean Lagrangian
\begin{equation}
Z=
\int_{\rm periodic} [d\theta]\exp -\int_0^\beta d\tau \int dV 
{\cal L}_{\rm E}(\theta, \mu)\, .
\label{eq3009}
\end{equation}
Since the exact functional integration over $\pi$ is  possible only 
if ${\cal H}$ is at most a quadratic function of $\pi$,
 we apply the saddle point approximation. 
In this approximation, the path integral is just the integrand 
evaluated for the field $\pi$ which solves
the saddle point condition
\begin{equation}
  i \frac{\partial\theta}{\partial\tau}
-\frac{\partial{\cal H}}{\partial\pi} +\frac{\mu}{m}=0.
\label{eq3010}
\end{equation}
Using (\ref{eq3008}) we find
\begin{equation}
\theta_{,0}=
  i \frac{\partial\theta}{\partial\tau}
 +\frac{\mu}{m}\, ,
\label{eq3011}
\end{equation}
 and with (\ref{eq3006}) we obtain
\begin{equation}
-{\cal L}_{\rm E}= i\pi \frac{\partial\theta}{\partial\tau}
 -\pi \theta_{,0}+\frac{\mu}{m}\pi
+{\cal L}(\theta_{,0},\theta_{,i})={\cal L}( i\frac{\partial\theta}{\partial\tau}
+\frac{\mu}{m},\theta_{,i}).
\label{eq3012}
\end{equation}
Hence, the effective Euclidean Lagrangian is obtained from
the Lagrangian (\ref{eq1203})
by replacing the derivatives of the field $\theta$ by
\begin{equation}
\theta_{,\nu}\rightarrow
\theta_{,\nu}+\frac{\mu}{m}\delta^0_\nu\, .
\label{eq3013}
\end{equation}
Note the difference and similarity with the usual Euclidean field theory
prescription \cite{kap}
\begin{equation}
\frac{\partial}{\partial \tau} \rightarrow
\frac{\partial}{\partial \tau}\pm \mu ,
\label{eq3203}
\end{equation}
for a canonical complex scalar field $\Phi$
where the $+$ or $-$ sign is taken when the derivative
acts on $\Phi^*$ or $\Phi$,
respectively.

\section{K-essence thermodynamics}
\label{thermodynamic}
In a grand canonical ensemble 
the thermodynamic quantities such as  pressure $p$, energy density $\rho$ and 
entropy density $s$ 
are functions of the temperature $T$ and  chemical potentials $\mu_i$ 
associated with conserved particle numbers $N_i$.
For simplicity, we consider only the case of a single
conserved particle number $N$ and its associated chemical potential
$\mu$.

We start from  the standard 
 thermodynamical equation 
\begin{equation}
sT=p+\rho-\mu n.
\label{eq1605}
\end{equation}
Clearly, if $\mu=0$ the positivity of
entropy requires $p+\rho\geq 0$.
Hence, ignoring $\mu$ one could conclude that a phantom field must necessarily
yield a fluid with negative entropy.
However, this conclusion is incorrect since
generally $\mu\neq 0$ and the entropy density given by
(\ref{eq1605}) need not be negative.
In fact, as we will shortly demonstrate, given arbitrary temperature $T$
it is always possible to find a range of $\mu$ such
that $s\geq 0$.

The entropy and 
particle-number  densities may be
expressed as partial derivatives of $p$
\begin{equation}
 s=\left.\frac{\partial p}{\partial T}\right|_\mu
\hspace{1cm}
n=\left.\frac{\partial p}{\partial \mu}\right|_T\, .
\label{eq1607}
\end{equation}
Using this and (\ref{eq1605}) we also find
\begin{equation}
 p+\rho=T\frac{\partial p}{\partial T}+
\mu \frac{\partial p}{\partial \mu}.
\label{eq1608}
\end{equation}

We now apply these general thermodynamic considerations to purely kinetic k-essence.
Equations (\ref{eq4003}) and (\ref{eq4004}) define a parametric EOS
which is essentially barotropic. In other words, by eliminating 
parametric dependence, the EOS 
may be put in the form $p=p(\rho)$.
Obviously, with a barotropic EOS
alone one can uniquely determine
neither $T$ nor $\mu$.
However, the k-essence relation
\begin{equation}
 p+\rho=2X{\cal L}_X \, ,
\label{eq1609}
\end{equation}
which follows from (\ref{eq4003}) and (\ref{eq4004}),
may be used to reduce the arbitrariness in
functional dependence on $T$ and $\mu$.
From (\ref{eq1608}) combined with (\ref{eq1609}) it follows that
the variable $X$ as a function of $T$ and $\mu$
satisfies a partial differential equation
\begin{equation}
 T\frac{\partial X}{\partial T}+
\mu \frac{\partial X}{\partial \mu}=2X.
\label{eq4005}
\end{equation}
The most general solution to this equation is 
a homogeneous function of 2nd degree
which may be written as
\begin{equation}
X= \frac{\mu^2}{m^2} f(T/\mu) .
\label{eq4006}
\end{equation}
Here $f$ is an arbitrary positive dimensionless function of $x\equiv T/\mu$.
However, the  consistency with
(\ref{eq3103}) places further restrictions on $f$.

The entropy and particle-number densities  may be calculated from $p$ using 
(\ref{eq4003}) and
(\ref{eq1607}).
With help of (\ref{eq4006}) we find
\begin{equation}
s= \frac{\mu}{m^2} f' {\cal L}_X; \hspace{1cm}
n=\left(\frac{2\mu}{m^2}f- \frac{T}{m^2}f'\right) {\cal L}_X .
\label{eq4009}
\end{equation}
Combining the second equation with (\ref{eq3103}) we obtain
a simple differential equation for $f$
\begin{equation}
xf'-2f+2\sqrt{f}=0 \, ,
\label{eq4108}
\end{equation}
with the solution
\begin{equation}
f=(Cx+1)^2\, ,
\label{eq4109}
\end{equation}
where $C$ is a constant.
Now we require $S=0$ at $T=0$,
which implies $f'(0)=0$ and hence $C=0$.
This in turn implies $f(x)=1$,
\begin{equation}
X=\frac{\mu^2}{m^2}\, ,
\label{eq411}
\end{equation}
 and $s=0$.
Hence, we come to the conclusion that
the  thermodynamic quantities, such as pressure and density,
derived from the classical kinetic k-essence field theory
are not temperature dependent and
can only depend on the chemical potential $\mu$.
Besides, we find that
the corresponding entropy is zero.
This result is not unexpected since it is well known that classical solutions to
a scalar  field theory (e.g., nontopological solitons) correspond to saddle point solutions
of the Euclidean path integral at nonzero $\mu$ and zero temperature \cite{lai,bil1}.
The existence of stable, nontopological solitons (Q-balls) that have a nonzero
value of the  conserved charge was proven \cite{col} for a complex canonical scalar field
theory of the type (\ref{eq1101}) for a class of ``acceptable'' potentials $U$.
Hence, based on the Thomas-Fermi equivalence mentioned in section \ref{pure}, 
we infer the existence of similar stable classical configurations 
in a corresponding class of purely kinetic k-essence.
Otherwise, if such configurations do not exist, then the only stable configuration 
at zero temperature is trivial, i.e., with $p=\rho=n=0$, in which case 
the entropy  $S$ is also zero.

However, if DE interacts with thermalized particles 
from the surroundings, its thermal fluctuations about the classical 
solutions would be in equilibrium with the heat bath at nonzero temperature.
In this case we will have $S>0$ as usual.
In  the next section we derive the corresponding thermal contribution to the
partition function.

\section{Thermal fluctuations}
\label{thermal}
In this section we derive the grand canonical thermodynamic potential
and canonical free energy for
a self-gravitating kinetic k-essence fluid
contained in a sphere of large radius
in equilibrium at nonzero temperature $T=1/\beta$.
Although the equilibrium assumption implies static metric, 
our analysis may be also applied to an expanding cosmology
in which case the thermodynamic parameters, such as 
temperature and chemical potential, are functions of
the scale parameter $a$. In this case 
the background metric $g_{\mu\nu}$ generated by the mass distribution is
assumed to be
a slowly varying function
of time on the inverse temperature scale.
More precisely, we assume
\begin{equation}
\frac{\partial g_{\mu\nu}}{\partial t} \ll \frac{g_{\mu\nu}}{\beta}\, .
\label{eq2005}
\end{equation}
We neglect the influence of  matter and radiation assuming that
their interaction with DE is small and serves only to provide a heat bath
at the temperature $T$.
\subsection{Grand canonical ensemble}
\label{grand}
In a grand canonical ensemble we introduce the chemical
 potential $\mu$ associated with the conserved particle
 number $N$ as in section \ref{chemical}.
The partition function is given by
\begin{equation}
Z= {\rm Tr}\, e^{-\beta(\hat{H}-\mu \hat{N})}=\int [dg][d\theta]
  e^{-S_{\rm g}-S_{\rm k}},
\label{z}
\end{equation}
with the Euclidean actions
 $S_{\rm g}$
and $S_{\rm k}$
 for the gravitational
and the k-essence fields, respectively.
The gravitational part may be put in the form
\cite{gib,haw}
\begin{equation}
S_{\rm g}=-\frac{1}{16\pi}\int_{Y} d^4x \sqrt{g_{\rm E}}  R
-\frac{1}{8\pi}\int_{\partial Y} d^3x \sqrt{h}(K-K_0),
\label{sg}
\end{equation}
where 
$h$ is the determinant of the induced metric on the boundary,
and $K-K_0$ is the difference
in the trace of the second fundamental form of the
boundary $\partial Y$ in the metric $g_E$ and the flat metric.
The boundary  is a timelike tube 
which is periodically identified in the imaginary time
direction with period $\beta$.
Thus, the functional integration assumes the periodicity
in imaginary time
and the asymptotic flatness of the metric fields.
The k-essence Euclidean action is 
\begin{equation}
S_{\rm k}=\int_Y d^4x \sqrt{g_{\rm E}}\,
{\cal L}_{\rm E}=-\int_Y d^4x \sqrt{-g}\,
{\cal L},
\label{skg}
\end{equation}
with the Euclidean Lagrangian
${\cal L}_{\rm E}$ given by (\ref{eq3012}).
The path integral is taken over
asymptotically vanishing fields
which are periodic
in imaginary time $\tau$ with
period $\beta$.
The dominant contribution to the path integral comes from metric
and k-essence fields, which are near the
classical  fields.
The  metric $g_{{\rm E}\mu\nu}$ is
assumed to be
static
spherically symmetric, and asymptotically flat,
and is positive
definite  with the Euclidean signature due to the
substitution (\ref{eq3105}).

In order to extract the classical contribution,
we decompose $\theta$ as
\begin{equation}
\theta(x)=\Theta(x)+m^{-2}\varphi(x),
\label{Phi}
\end{equation}
where $\Theta$ is a solution to the
classical equation of motion
which we will call {\em condensate}
and $\varphi$ describes  quantum and thermal
fluctuations around  $\Theta$.
 The factor $m^{-2}$ is introduced 
for convenience so that 
the field $\varphi$ has the 
canonical dimension one.
The action $S_{\rm k}$ splits in two parts
$S_{\rm k}=S_{\rm cl}[\Theta]+S_{\rm th}[\varphi]$
and if we neglect quantum fluctuations
of the metric, the partition function factorizes as
\begin{equation}
Z=Z_{\rm g} Z_{\rm cl}
Z_{\rm th}\, ,
\label{zz}
\end{equation}
where $Z_{\rm g}=e^{-S_{\rm g}}$ and
$Z_{\rm cl}=e^{-S_{\rm cl}}$ represent the saddle-point
 gravitational and classical contributions, respectively.
 The last factor is the contribution due to thermal fluctuations
 \begin{equation}
Z_{\rm th}=
\int [d\varphi]e^{-S_{\rm th}[\varphi]}.
\label{zth}
\end{equation}

First, we calculate the classical contribution.
The classical part of the action
is given by
\begin{equation}
S_{\rm cl}=-\int_Y d^4x \sqrt{-g}\:
{\cal L}(X).
\label{s0}
\end{equation}
Here and below
\begin{equation}
X=g^{\mu\nu}\Theta_{,\mu}\Theta_{,\nu}\,,
\label{x}
\end{equation}
where
\begin{equation}
\Theta_{,0}=i\frac{\partial\Theta}{\partial \tau}+\frac{\mu}{m}=
\frac{\partial\Theta}{\partial t}+\frac{\mu}{m}\,,
\label{xx}
\end{equation}
\begin{equation}
\Theta_{,i}=\frac{\partial\Theta}{\partial x_i}\, .
\label{xxx}
\end{equation}
By
 making  use of 
the field equation 
 (\ref{eq1304}) in the comoving frame, i.e., in the reference frame
 where the 4-velocity components are
\begin{equation}
u^{\mu}=
\frac{
\delta^{\mu}_0
}{\sqrt{g_{00}}}\,  ;
 \;\;\;\;\;\;\;
u_{\mu}=
\frac{
g_{\mu 0}
}{\sqrt{g_{00}}}\,,
\label{u}
\end{equation}
it follows
\begin{equation}
\Theta =\omega t + \vartheta=-i\omega \tau+ \vartheta\, ,
\label{eq2202}
\end{equation}
where $\omega$ and $\vartheta$ are constants.
Furthermore, using (\ref{x}) with (\ref{xx}) and comparing with (\ref{eq411})
we conclude that $\omega=0$. Thus, the background field $\Theta$ is constant.
However, since $\Theta_{,0}$ is given by
(\ref{xx}), the quantity $X$  may still
 be a function of
$\vec{x}$ and   a slowly varying function of time.
Equations (\ref{x})-(\ref{xx}) give
\begin{equation}
X= g^{00}\frac{\mu^2}{m^2}=\frac{\bar{\mu}^2}{m^2}\, ,
\label{eq2203}
\end{equation}
where $\bar{\mu}=\mu/\sqrt{g_{00}}$ is the local chemical potential.
This equation is consistent with equation (\ref{eq411}) obtained from purely 
thermodynamic considerations.

In a spherically
symmetric  static
geometry,
regular solutions to
equation (\ref{eq1304}) coupled with Einstein field equations
 describe
dark energy stars
\cite{chap,lobo} at zero temperature. A particular example of Born-Infeld type
k-essence stars have recently been studied in \cite{bil8}.
 In the context of cosmology, the background is
spatially homogeneous isotropic configuration with $X$ depending  on
time through cosmological scale dependence.

According to our assumption (\ref{eq2005})
the variation of $X$ is small on the time scale comparable with  
the inverse temperature, i.e., $\partial X/\partial t \ll X/\beta$. 
Then, the condensate
contribution to the partition function may be written as
\begin{equation}
\ln Z_{\rm cl}=\beta
\int_{\Sigma} d^3x \sqrt{-g}\,
{\cal L} (X)
\label{lnz0}
\end{equation}
where $\Sigma$ is a spacelike hyper-surface
that contains the condensate.
Using this equation
we find the net number of particles in the condensate
\begin{equation}
N=\frac{1}{\beta}\frac{\partial \ln Z_{\rm cl}}{
\partial \mu}
=\int_{\Sigma} d^3x \sqrt{g_{(3)}}\:
\frac{2\bar{\mu}}{m^2}{\cal L}_X 
\label{n0}
\end{equation}
where $g_{(3)}=\det (-g_{ij})$, $i,j=1,2,3$.
Alternatively, the covariant definition (\ref{eq3002}) yields
\begin{equation}
N=
\int_{\Sigma} d^3x \sqrt{g_{(3)}}\: n \,,
\label{n0s}
\end{equation}
where
$n$ is the particle-number density
in the condensate.
Therefore, we identify the  particle-number density
due to the condensate as
\begin{equation}
n=
2\frac{\bar{\mu}}{m^2}{\cal L}_X 
\label{ncd}
\end{equation}
which coincides with
(\ref{eq3103}) as it should.

Next, we calculate the thermal
contribution to $Z$ starting from 
equation
(\ref{zth})
where the action $S_{\rm th}[\varphi]$ is derived from ${\cal L}(X)$
by keeping only the quadratic term
in the expansion
of ${\cal L}$ in powers of $\varphi$. 
We find
\begin{equation}
S_{\rm th} =
-m^{-4}\int_Y d^4x \sqrt{-g}\,
f^{\mu\nu}\partial_{\mu}\varphi
\partial_{\nu}\varphi \, ,
\label{skg1}
\end{equation}
where 
\begin{equation}
f^{\mu\nu}={\cal L}_X g^{\mu\nu}+
2{\cal L}_{XX}g^{\mu\alpha}g^{\mu\beta}\Theta_{,\alpha} \Theta_{,\beta}\, .
\label{eq2006}
\end{equation}
The symbol $\partial_\mu$ in equation (\ref{skg1})  denotes
the partial derivative with respect to a Lorentzian coordinate $x_\mu$.
In particular, $\partial_0=\partial/\partial t=i\partial/\partial \tau$.
The $\mu$ dependence is absorbed in $f^{\mu\nu}$ through the prescription
(\ref{xx}).
The action describes a massless scalar propagating in 
an effective (or emergent) acoustic
geometry \cite{mon,bilCQG,arm3,bab} provided the field equation for $\varphi$
\begin{equation}
f^{\mu\nu}\varphi_{;\mu\nu}=0
\label{eq2014}
\end{equation}
is hyperbolic, i.e., provided the effective metric tensor $f^{\mu\nu}$ has the Lorentzian signature.
This holds if and only if
\begin{equation}
\det f^{\mu\nu}={\cal L}_X^4 g^{-1}c_s^{-2} < 0,
\label{eq2015}
\end{equation}
where 
\begin{equation}
c_s^{-2}=\frac{{\cal L}_X+2{\cal L}_{XX}}{{\cal L}_X} 
\label{eq2011}
\end{equation}  
is the inverse speed of sound squared.
It may be easily verified that the same
 equation is obtained from the standard 
hydrodynamic definition
  \cite{lan}
\begin{equation}
c_s^2=\left. \frac{\partial p}{\partial\rho}
 \right|_{s/n} .
\label{eq016}
\end{equation}
Since $g<0$, $f^{\mu\nu}$ is Lorentzian if and only if
$c_s^2>0$. Hence, the hyperbolicity condition (\ref{eq2015})
is equivalent to the requirement of hydrodynamic stability.

In addition, it seems physically reasonable to require 
$c_s^2<1$ in order to avoid possible problems with causality.
However, it has been shown \cite{bon,kan} that if k-essence is to solve
the coincidence problem there must be 
an epoch when perturbations in the k-essence field propagate faster than light.
In cosmology and astrophysics, it is usually assumed on the basis of causality 
that the speed of sound cannot exceed the
speed of light \cite{ell,wal,ell2}.
Hence, it has been argued \cite{bon}
that  k-essence models which 
solve the coincidence problem are ruled out as 
realistic physical candidates for dark energy.
In contrast to this, it has been argued \cite{bab,kan,bru} that 
superluminal sound speed propagation in generic k-essence models does not 
necessarily lead to
causality violation and hence,
 in spite of the presence
of superluminal signals on nontrivial backgrounds, the k-essence theories are not less 
legitimate than General Relativity.

As we are not presently interested in discussing or solving the coincidence problem
we stick to $0<c_s^2<1$. This condition is fulfilled if
both ${\cal L}_X$ and ${\cal L}_{XX}$ are simultaneously
either positive or negative. In the latter case
from (\ref{eq1609}) we have $p+\rho<0$ and hence, our thermodynamic analysis 
is not restricted only to 
models that satisfy the null energy condition.

Assuming $0<c_s^2<1$,
 equation (\ref{skg1})
may be put in the form
\begin{equation}
S_{\rm th} =
-\frac{1}{2}\int_Y d^4x \sqrt{-G}\,
G^{\mu\nu}\partial_{\mu}\varphi
\partial_{\nu}\varphi \, .
\label{eq2007}
\end{equation}
where
\begin{equation}
G^{\mu\nu}=2\frac{c_s m^4}{{\cal L}_X}
[g^{\mu\nu}-(1-\frac{1}{c_s^2})u^\mu u^\nu]\, .
\label{eq2208}
\end{equation}
The matrix $G^{\mu\nu}$ is the inverse of
the acoustic metric tensor defined as \cite{mon,bilCQG}
\begin{equation}
G_{\mu\nu}=\frac{1}{2}\frac{{\cal L}_X}{c_s m^4}
[g_{\mu\nu}-(1-c_s^2)u_\mu u_\nu]\, ,
\label{eq2008}
\end{equation}
where
\begin{equation}
u_\mu=\frac{\Theta_{,\mu}}{\sqrt{X}}\,; \hspace{1cm}
u^\mu= g^{\mu\nu}u_\mu
\label{eq2009}
\end{equation}
is the velocity of the fluid.

The determinant $G$ is given by
\begin{equation}
G=\det G_{\mu\nu}=\frac{16{\cal L}_X^4}{m^{16} c_s^2}g .
\label{eq2010}
\end{equation}
Since the metric is static by assumption , i.e.,
$g_{\mu\nu}$ is independent of $t$ and
$g_{0 i}=0$, the same is true for the acoustic metric
$G_{\mu\nu}$ and the  determinant  factorizes
as
$G=-G_{00}
G_{(3)}$ where
$
G_{(3)}=-\det G_{ij}$; $i,j=1,2,3$.
By making use of the substitution
$\tilde{\tau}=\tau \sqrt{G_{00}}$
we obtain
\begin{equation}
S_{\rm th} =
\int_{\Sigma} d^3x \sqrt{G_{(3)}}
\int_0^{\tilde{\beta}}d\tilde{\tau}
{\cal L}(\tilde{\tau},x) ,
\label{skg2}
\end{equation}
where
\begin{equation}
{\cal L}(\tilde{\tau},x) =\frac{1}{2}
[(\partial_{\tilde{\tau}} \varphi)^2
 -G^{ij}\partial_i\varphi
\partial_j\varphi]\,.
 \label{l}
\end{equation}
Here we have introduced the effective inverse temperature
\begin{equation}
\tilde{\beta}=\sqrt{G_{00}}\beta=\sqrt{\frac{G_{00}}{g_{00}}}\bar{\beta}\, ,
\label{eq2012}
\end{equation}
where the quantity $\bar{\beta}=\sqrt{g_{00}}\beta$ is the
usual local inverse temperature.
Equation(\ref{eq2012}) is
 the well-known Tolman condition for 
thermal  equilibrium in curved space
\cite{tol,lan}.
The parameter $\beta$  retains its usual interpretation as the asymptotic  value
of the inverse temperature \cite{bar}.

Using the approach developed in \cite{bil1} 
the path integral in (\ref{zth}) may be easily
calculated. 
We obtain the thermal part of the partition function in the form 
\begin{equation}
\ln Z_{\rm th}=
-\int_\Sigma d^3x\,\sqrt{G_{(3)}}
 \int \frac{d^3q}{(2\pi)^3}
 \ln (1-e^{-\tilde{\beta}q}) .
 \label{lnzth}
\end{equation}
This expression may be regarded as a proper volume integral
\begin{equation}
\ln Z_{\rm th}=
-\int_\Sigma d^3x\,\sqrt{g_{(3)}}\:
 \frac{1}{V} \ln z 
 \label{lnzth2}
\end{equation}
of the local partition function
\begin{equation}
\ln z =
- V\sqrt{\frac{G_{(3)}}{g_{(3)}}}\int \frac{d^3q}{(2\pi)^3}
 \ln (1-e^{-\tilde{\beta}q})=
 \frac{V\pi}{360}\sqrt{\frac{G_{(3)}}{g_{(3)}}}
\tilde{\beta}^{-3},
 \label{lnz}
\end{equation}
from which
the pressure,
energy density,
and entropy density may be derived in the usual way:
\begin{equation}
p_{\rm th}=
\frac{1}{V\bar{\beta}} \ln z=
\frac{\pi}{360}c_s^{-3}\bar{\beta}^{-4},
 \label{pth}
\end{equation}
\begin{equation}
\rho_{\rm th}=
-\frac{1}{V}\frac{\partial}{\partial
\bar{\beta}}
\ln z =
 \frac{\pi}{120}c_s^{-3}\bar{\beta}^{-4},
 \label{rhoth}
\end{equation}
\begin{equation}
s=
\bar{\beta}
(p_{\rm th} +
\rho_{\rm th})= 
\frac{\pi}{90}c_s^{-3}\bar{\beta}^{-3}.
 \label{sigma}
\end{equation}
 These expressions are
 the well-known thermodynamic 
equations that represent a massless relativistic Bose
gas in curved space. The factor $c_s^{-3}$ is the effective number of degrees
of freedom which depends on the chemical potential 
for a particular k-essence model. 
This shows that a sizable fraction
of radiation today can be attributed to thermal
fluctuations of DE.
For example, for the Born-Infeld type k-essence
(\ref{eq0070}) we find 
\begin{equation}
c_s^{-3}=(1-\mu^2/m^2)^{-3/2}.
 \label{eq2013}
\end{equation}
If we use this
model for a dark matter/energy unification,
 the fit with homogeneous cosmology today would yield
the speed of sound squared of the order $c_s^2=\Omega_\Lambda$
with the effective number of degrees of freedom $\Omega^{-3/2}\approx 1.6$,
comparable with $2$ for photons.

The gravitational part of the partition function
may be calculated
from (\ref{sg}) with help of Einstein field equations.
Using the result of Gibbons and Hawking
for the surface term
\cite{gib}, we obtain
\begin{equation}
\ln Z_{\rm g}=-\beta M+\int_Y d^4x \sqrt{g} \,
T_0^0  ,
\label{lnzg}
\end{equation}
where $M$ is the total mass and $T_{\mu}^{\nu}$ is
the energy-momentum tensor of k-essence field
averaged with respect to the partition function (\ref{zz}).
The averaged energy-momentum tensor may be split up into
two parts:
\begin{equation}
 T_{\mu\nu}=
 T_{{\rm cd}\:\mu\nu}+
 T_{{\rm th}\:\mu\nu},
\label{tmunu}
\end{equation}
where the first term on the right-hand side is
the classical part which comes from
the condensate and the
 second term
 represents
 the thermal  fluctuations.

 It may be shown that both terms are of the 
form (\ref{eq510})
 which characterizes a perfect fluid.
Hence, we have 
\begin{equation}
 T_0^0=
\rho_{\rm cd}+\rho_{\rm th} \, ;
\hspace{1cm}
T_i^i=
p_{\rm cd}+p_{\rm th} \,,
\label{tt}
\end{equation}
where the 
thermal pressure and
 energy density
are given by
 (\ref{pth}) and (\ref{rhoth}),
respectively. 
The condensate pressure $p_{\rm cd}$ and energy density $\rho_{\rm cd}$
are given by (\ref{eq4003}) and (\ref{eq4004}), respectively, in which 
the quantity $X$ is given  by (\ref{eq2203}).
Putting  the condensate (\ref{lnz0}),
the thermal (\ref{lnzth2}), and
the gravitational (\ref{lnzg})
contributions together,
we find the total grand canonical thermodynamic potential as
\begin{equation}
\Omega (\beta, \mu) \equiv
-\frac{1}{\beta} \ln Z =
M-
\int_{\Sigma} d^3x \sqrt{-g}
\left( p_{\rm cd}
+\rho_{\rm cd}
+p_{\rm th}
+\rho_{\rm th}
 \right).
\label{omega}
\end{equation}
This is just the standard form  of
the thermodynamic potential 
\begin{equation}
\Omega=
E-TS-\mu N,
\label{omega2}
\end{equation}
in which we identify the energy $E$ with the total mass $M$,
the $TS$ term with the 
thermal contribution
\begin{equation}
TS=
\int_{\Sigma} d^3x \sqrt{-g}
\left( 
p_{\rm th}
+\rho_{\rm th}
 \right),
\label{entro}
\end{equation}
and the $\mu N$ term with the contribution of the condensate
\begin{equation}
\mu N=
\int_{\Sigma} d^3x \sqrt{-g}
\left( p_{\rm cd}
+\rho_{\rm cd}
 \right)
=\mu \int_{\Sigma} d^3x \sqrt{g_{(3)}}\, n .
\label{particle}
\end{equation}
Note that the entropy $S$  
 is strictly positive because the right-hand side of 
(\ref{entro}) is positive and the quantity $T$
is a positive temperature of the heat bath.
\subsection{Canonical ensemble}
Now consider
a  self-gravitating k-essence fluid
with $N$ particles
contained in a two-di\-men\-sio\-nal sphere of large radius 
in equilibrium at nonzero temperature $T=1/\beta$.
In a
 canonical ensemble, instead of the chemical
 potential $\mu$ we fix  the particle number. 
Thus, a canonical ensemble is subject to
the constraint (\ref{n0s}) with $N$ fixed.

The free energy of a canonical ensemble may
be derived from the grand canonical partition function
with the help of the Legendre transform
\begin{equation}
F(\beta,N)=
\Omega (\beta, \mu) + \mu N.
\label{f}
\end{equation}
 The quantity $\mu$
 in this expression
 is an implicit
 function of $N$ and $T$, such that for given $N$ and $T$
 the constraint (\ref{n0s}) is satisfied
 {\cite{bilGRG}}.
 From (\ref{omega}) and (\ref{n0s}) with (\ref{ncd}) it follows that
\begin{equation}
F =
M-
\int_{\Sigma} d^3x \sqrt{-g}\:
(p_{\rm th}
+\rho_{\rm th}).
\label{fm}
\end{equation}
The second term on the right-hand side is related to the entropy
through (\ref{sigma})
and the free energy may be expressed in the familiar form
\begin{equation}
F=
M-TS,
\label{fmt}
\end{equation}
where the total entropy $S$ is defined as a proper volume
integral
\begin{equation}
S=
\int_{\Sigma} s
u^{\mu}d\Sigma_{\mu}
=\int_{\Sigma} d^3x \sqrt{g_{(3)}}\, s
\label{s}
\end{equation}
of the entropy density $s$ given by
(\ref{sigma}).

\subsection{More general k-essence}
\label{general}
The above considerations may be similarly applied to 
a general class of  k-essence models 
described by
\begin{equation}
S = \int \, d^{4}x \, \sqrt{- g}  \left[ - \frac{R}{16\pi G} + {\cal L} (\theta,X) \right],
\label{eq5001}
\end{equation}
with the most general Lagrangian 
which, in addition to $X$,  depends explicitly
on the scalar field
$\theta$.
The hydrodynamic quantities $u_\mu$, $p$, $\rho$, $T_{\mu\nu}$,
and $c_s$  associated with (\ref{eq5001}) are, 
as before, defined by (\ref{eq1215})-(\ref{eq4010}).
However, unlike in purely kinetic k-essence,
the equation of motion for $\theta$
\begin{equation}
(2{\cal L}_X g^{\mu \nu} \theta_{, \nu})_{;\mu}-\frac{\partial\cal L}{\partial\theta}=0
\label{eq2304}
\end{equation}
is no longer of the current conservation form
and the analysis of sections  
\ref{chemical} and \ref{thermodynamic} 
does not  apply.
Here, for example, we cannot establish a correspondence with the
canonical complex scalar field theory.
As in the case of a real scalar field, 
 because of the absence of a conserved Noether current
there exist no stable nontopological solutions
\cite{lee} although
time dependent solutions
similar to oscillatons \cite{sei,ure}, or static unstable configurations
similar to unstable scalar solitons \cite{alc}, are not excluded.
The only stable solutions at zero temperature
are trivial, i.e., those with $p=\rho=0$ so that 
the condensate contribution to the partition function is absent.

As there is no conserved particle number the canonical and grand canonical
ensembles coincide, i.e., $F(\beta)=\Omega(\beta)$, 
and the free energy $F$ is given by 
(\ref{fm}).
The thermal contribution is of the form (\ref{zth}) 
where the action $S_{\rm th}[\varphi]$ is
obtained by expanding 
${\cal L}(X,\theta)$ in powers of $\varphi$
with $\theta=\Theta+m^{-2}\varphi$, as before.
Here, $\Theta$ describes a configuration for which
$p=\rho=0$.
Using the procedure described in section \ref{grand}
we find the expression similar to (\ref{eq2007})
\begin{equation}
S_{\rm th} =
-\frac{1}{2}\int_Y d^4x \sqrt{-G}\,
(G^{\mu\nu}\partial_{\mu}\varphi
\partial_{\nu}\varphi -m_{\rm eff}^2\varphi^2)\, ,
\label{eq4007}
\end{equation}
where
\begin{equation}
m_{\rm eff}^2= \frac{m^4c_s}{4{\cal L}^2_X}\left(2X{\cal L}_{X\Theta\Theta} -
{\cal L}_{\Theta\Theta}+2\frac{\partial f^{\mu\nu}}{\partial\Theta}\Theta_{;\mu\nu}\right)
\label{eq4008}
\end{equation}
is the effective mass \cite{bab},
$X=g^{\mu\nu}\Theta_{,\mu}\Theta_{,\nu}$,
and the matrices $f^{\mu\nu}$ and $G^{\mu\nu}$ are defined as in 
(\ref{eq2006}) and (\ref{eq2208}), respectively.
The subscripts $X$ and $\Theta$  in (\ref{eq4008}) denote 
the partial derivatives
 with 
respect to $X$ and $\Theta$, respectively.
The quantity $c_s$ is the sound speed defined as in (\ref{eq2011}).
The action (\ref{eq4007})
describes a massive scalar propagating in 
an effective acoustic
geometry provided the condition
(\ref{eq2015})
is met. Again, using the approach developed in \cite{bil1}
we find the expression for the thermal partition function
in the form of the proper volume integral (\ref{lnzth2})
 over
the local partition function
\begin{equation}
\ln z =
- V\sqrt{\frac{G_{(3)}}{g_{(3)}}}\int \frac{d^3q}{(2\pi)^3}
 \ln (1-e^{-\tilde{\beta}E})
 \label{lnz2}
\end{equation}
where $E=\sqrt{m_{\rm eff}^2+q^2}$ and $g_{(3)}$ and $G_{(3)}$
are the determinants of the respective spatial metrics
defined as in section \ref{grand}.
 From this partition function
we find the usual  expressions for the
 pressure, energy density, and entropy density of an
ideal gas of massive bosons
\begin{equation}
p_{\rm th}=\frac{g_{\rm eff}}{\bar{\beta}}
\int \frac{d^3q}{(2\pi)^3}
 \ln (1-e^{-\bar{\beta}E}),
 \label{pth2}
\end{equation}
\begin{equation}
\rho_{\rm th}=g_{\rm eff}
\int \frac{d^3q}{(2\pi)^3}
 \frac{1}{1-e^{-\bar{\beta}E}} \, ,
 \label{rhoth2}
\end{equation}
\begin{equation}
s=\bar{\beta}(p_{\rm th}+\rho_{\rm th})
 \label{entr2}
\end{equation}
with the effective number of degrees of freedom 
$g_{\rm eff}=1/c_s^3$  which depends on the model.
\section{Conclusions}
\label{conclusion}

We have derived
a grand canonical and canonical description of 
k-essence type of DE.
The thermodynamic equations are 
generally expressed in terms  of two variables:
the temperature $T$ and chemical potential $\mu$.
The chemical potential is associated
with a conserved particle number  $N$
related to the shift symmetry.
The derived thermodynamic equations show
that the entropy does not have to be negative even in the phantom regime,
contrary to the claims
often stated in the recent literature
(see, e.g., \cite{gong} and references therein)
that a violation of
the null energy condition  implies
negative entropy.
We have demonstrated that the entropy is greater or equal to zero
and is strictly zero at zero temperature.
We have show that if there exist a nontrivial, stable configuration
which we call {\em condensate}
characterized by the pressure $p_{\rm cd}$ and the density $\rho_{\rm cd}$,
then the particle-number term in the expression for entropy cancels out
the contribution of $p_{\rm cd}$ and $\rho_{\rm cd}$. 
Furthermore, we have shown that the  only nonvanishing
contribution to the entropy is due to thermal fluctuation
analogous to those of a massless boson field.
The  thermal ensemble behaves as a free massless gas
at nonzero temperature with an effective number of degrees of freedom equal 
to $c_s^{-3}$. Similarly, thermal fluctuations of a general k-essence field
yield an effective free gas of massive bosons.


\subsection*{Acknowledgments}
This work is supported by the Ministry of Science,
Education and Sport
of the Republic of Croatia under contract No. 098-0982930-2864 and
partially supported through the Agreement between the Astrophysical
Sector, S.I.S.S.A., and the Particle Physics and Cosmology Group, RBI.

\end{document}